\documentclass[twocolumn,aps,pra,showpacs,bibfootnote,preprintnumbers, mathtools, superscriptaddress, amsmath,amssymb]{revtex4-2}
\usepackage[utf8]{inputenc}
\usepackage[english]{babel}
\usepackage[T1]{fontenc}
\usepackage{amsmath}
\usepackage{hyperref}
\usepackage{float}
\usepackage{orcidlink}

\usepackage{tikz}
\usepackage{lipsum}

\usepackage{graphicx}
\usepackage{dcolumn}
\usepackage{bm}
\usepackage{mathtools}          
\usepackage{braket}
\usepackage{color}
\usepackage[final]{changes}
\usepackage{braket}
\usepackage{xcolor}
\usepackage{amsmath}
\usepackage{accents}
\usepackage{graphicx,caption}

\newcommand{\lBopp}[1]{\accentset{\curvearrowright}{#1}}
\newcommand{\rBopp}[1]{\accentset{\curvearrowleft}{#1}}

\newcommand{\elBopp}[1]{\expandafter\lBopp#1}
\newcommand{\erBopp}[1]{\expandafter\rBopp#1}
\newcommand{\Tr}{\mathrm{Tr}}
\newcommand{\bi}[1]{\mathbf{#1}}

\begin{document}

\title[The waveoperator representation of quantum and classical dynamics]{The waveoperator representation of quantum and classical dynamics}

\author{Gerard McCaul   }%
 \email{gmccaul@tulane.edu}
 \address{Tulane University, New Orleans, LA 70118, USA}
 \author{Dmitry V. Zhdanov}%
 \address{Tulane University, New Orleans, LA 70118, USA}
 \author{Denys I. Bondar \orcidlink{0000-0002-3626-4804}}%
 \address{Tulane University, New Orleans, LA 70118, USA}

\begin{abstract}
The choice of mathematical representation when describing physical systems is of great consequence, and this choice is usually determined by the properties of the problem at hand. Here we examine the little-known waveoperator representation of quantum dynamics, and explore its connection to standard methods of quantum dynamics, such as the Wigner phase space function. This method takes as its central object the \textit{square root} of the density matrix, and consequently enjoys several unusual advantages over standard representations. By combining this with purification techniques imported from quantum information, we are able to obtain a number of results. Not only is this formalism able to provide a natural bridge between phase and Hilbert space representations of both quantum and classical dynamics,  we also find the waveoperator representation leads to novel semiclassical approximations of both real and imaginary time dynamics, as well as a transparent correspondence to the classical limit. {It is then demonstrated that there exist a number of scenarios (such as thermalisation) in which the waveoperator representation possesses an equivalent unitary evolution, which corresponds to \textit{non-linear} real-time dynamics for the density matrix}. We argue that the waveoperator provides a new perspective that links previously unrelated representations, and is a natural candidate model for scenarios (such as hybrids) in which positivity cannot be otherwise guaranteed. 
\end{abstract}
\maketitle
\section{Introduction}
When describing physical dynamics mathematically, there exist a number of equivalent representations that one can choose between. This plethora of potential representations is particularly pronounced in quantum dynamics. Besides the Schr{\"o}dinger and Liouville equations, there also exists more esoteric formulations such as the Wigner-Weyl phase space representation \cite{doi:10.1119/1.2957889,PhysRev.40.749, Baker1958,curtright_concise_2014,Groenewold1946}, or the Feynman path integral \cite{Schulman2005-av}. Each of these carries its own strengths and weaknesses. For example, the phase space representation is commonly used in quantum chemistry and optics \cite{curtright_concise_2014}, while path integrals find a natural home in the description of open system dynamics via the influence functional \cite{Feynman-Vernon-1963,PhysRevB.95.125124,PhysRevB.97.224310,McCaul2021,Caldeira1983,Ford-Kac-JST-1987,Gardiner-1988,Sebastian1981,Leggett1987,van_Kampen-1997}. On a more fundamental level, the choice of representation can change the degree which the correspondence principle is manifestly present. To draw again on the example of the path integral and Wigner function,  the $\hbar \to 0$ limit makes clear that for the former the only path of finite weight is that corresponding to the classical action \cite{McCaul2021}, while the equation of motion for the Wigner function reduces to the classical Poisson bracket.

In the realm of quantum dynamics, one's choice of representation can often lead to issues of interpretation. For instance, the measure of a path integral is only finitely additive and therefore not guaranteed to converge \cite{Klauder2003}, while the Wigner function exhibits negativity. This is particularly problematic, as this potential negativity means it is uninterpretable as a density, despite being derived from one. In the case of pure states, this difficulty was resolved with the demonstration that the Wigner function should be interpreted as a phase space probability amplitude~\cite{bondar_wigner_2013}. This is in direct analogy with the Koopman-von Neumann (KvN) representation of classical dynamics \cite{McCaul_2022,PhysRevE.99.062121, Bondar2019, Koopman315, Wilkie1997a,Wilkie1997b,Sudarshan1976,Viennot_2018,PhysRevLett.120.070402,SEN2020168302,DHASMANA2021168623,SEN2023104732, BermdezManjarres2021,BermdezManjarres2022} which explicitly admits a wavefunction on phase space, and which the Wigner function of a pure state corresponds to in the classical limit. The extension of this interpretation to mixed states has to date been lacking however, given that such states must be described by densities and therefore lack a direct correspondence to wavefunctions. 

Here we address this issue, by employing the little-known \textit{waveoperator} formalism.  Such a representation of dynamics has been deployed in a number of contexts, including open systems \cite{ilya}, quantum holonomy in density matrices \cite{uhlmann_parallel_1986}, the development of phase-space wavefunctions \cite{wlodarz_quantum_1994}, as well as nonlinear dynamical models \cite{gheorghiu-svirschevski_nonlinear_2001, beretta_nonlinear_2006, beretta_nonlinear_2005}. In fact, the motivation for a ``square root'' of the density and the advantages it provides can be found even when not explicitly referenced. For example, the recently developed ensemble rank truncation~\cite{McCaulensemble} and corner-space renormalization~\cite{le_bris_low-rank_2013, finazzi_corner-space_2015, donatella_continuous-time_2021} methods have a their heart a method for representing a Lindbladian evolution of a density in terms of a weighted sum of wavefunctions. The waveoperator has also been used for foundational research~\cite{reznik_unitary_1996, yahalom_square-root_2006, tronci_momentum_2019}, but here we extend this to demonstrate that when combined with purification techniques from quantum information, it provides a natural bridge between the Hilbert space representation of quantum dynamics, the phase space Wigner representation of the same, as well as KvN dynamics \cite{mezic2005spectral}. Through this, we are able to derive not only a consistent interpretation of mixed states in the Wigner representation, but also establish a connection between the commonly utilised phase space methods of quantum chemistry, and quantum information. We also find that this representation of quantum dynamics leads to novel semiclassical approximations of both real and imaginary time dynamics, as well as a transparent correspondence to the classical limit. {In the context of thermalising imaginary time evolutions, the waveoperator representation is also be used to establish that such dynamics may be directly mapped to unitary real-time dynamics which are explicitly \textit{non-linear} on the level of densities.} 
 
 The remainder of this paper is outlined as follows - Sec.~\ref{Sec:purification} borrows from the field of quantum information to express the waveoperator in a purified form. This is then exploited in Sec.~\ref{SecBoppOp} to introduce Bopp operators into the waveoperator description. Equipped with this formulation, in Sec.~\ref{Sec:phasespace} it is possible to identify the phase-space representation of the waveoperator with the Wigner function, while in Sec.~\ref{SecClassicalLimit} we use it to demonstrate that the classical limit of the waveoperator description reduces exactly to the KvN representation of classical dynamics. Sec.~\ref{Sec:Imaginarytime} then applies the same technique to the imaginary time Bloch equation, where we are able to derive a semi-classical correction to the equilibrium state of a system, and illustrate its effect using the examples of a quadratic and quartic oscillator. { Sec.~\ref{sec:nonlinearity} demonstrates two further scenarios, one in which linear waveoperator dynamics leads to nonlinear dynamics in the density matrix, and a complementary case where the opposite relationship occurs.} The paper then concludes with a summary of key findings, as well as outlining both open questions and future research directions.

\section{ The waveoperator \label{Sec:waveoperator}}

We begin our treatment by making explicit a freedom present in the in the Liouville equation. This describes the dynamics for the density matrix $\hat{\rho}$ of a quantum system: 
\begin{equation}\label{EqQuantumLiouvilleEq}
    i\hbar \partial_t \hat{\rho} = [\hat{H}, \hat{\rho}] 
    \equiv  \hat{H} \hat{\rho} - \hat{\rho} \hat{H}, 
\end{equation}
where $\hat{H}$ is a self-adjoint Hamiltonian. The expectation value of an observable $\hat{O}$ is obtained as
\begin{equation}\label{EqRhoAverage}
    \braket{O} = \Tr(\hat{\rho}\hat{O}).
\end{equation}
{First, the density matrix, as a non-negative operator, can always be decomposed into the form}: 
\begin{equation}\label{EqRhoThroughOmega}
    \hat{\rho} = \hat{\Omega} \hat{\Omega}^\dagger,
\end{equation}
where in what follows we shall refer to $\hat{\Omega}(t)$ as the \textit{waveoperator}. Following this assignation, we might ask what form the dynamics of $\hat{\Omega}(t)$ can take while remaining consistent with both Eqs.~\eqref{EqQuantumLiouvilleEq} and~\eqref{EqRhoThroughOmega}. We find that the most general form of evolution permitted is
\begin{equation}\label{EqWaveOperatorDynamics}
    i\hbar \partial_t \hat{\Omega} = [\hat{H}, \hat{\Omega}] -  \hat{\Omega}\hat{F},
\end{equation}
where $\hat{F}$ is an arbitrary self-adjoint operator. It is easy to show from this that
\begin{equation}\label{EqDensityMatrixDynamics}
    i\hbar \partial_t ( \hat{\Omega}\hat{\Omega}^{\dagger} ) = [\hat{H}, \hat{\Omega}\hat{\Omega}^{\dagger} ]
\end{equation}
is satisfied at all times if it is satisfied at a single moment (e.g., $t=0$). Consequently, the Liouville dynamics described by Eq.~\eqref{EqQuantumLiouvilleEq} may instead be described via the waveoperator using Eq.~\eqref{EqWaveOperatorDynamics}, together with a prescription for expectations:
\begin{equation}\label{EqOAverageWaveOperator}
    \braket{O} = \Tr( \hat{\Omega}^\dagger \hat{O} \hat{\Omega}).
\end{equation}

The principal advantages of expressing a quantum system's dynamics in terms of $\hat{\Omega}(t)$ rather than $\hat{\rho}(t)$ are two-fold. {First, by evolving the ``square root" of the density, any dynamics using $\hat{\Omega}(t)$ are guaranteed to enforce the positivity of $\hat{\rho}$. While in the forms of dynamics considered in the present paper this positivity is already preserved for $\hat{\rho}$, it is a significant benefit when considering scenarios such as quantum chemistry applications~\cite{DMM}, open quantum systems~\cite{ilya}, and quantum-classical hybrids~\cite{aleksandrov_statistical_1981, gerasimenko_dynamical_1982}, where the dynamics of $\hat{\rho}$  do not identically preserve this property. }

{The presence of $\hat{F}$ in Eq.\eqref{EqWaveOperatorDynamics} is a direct consequence of this automatic positivity preservation. The physical meaning of this addition can be understood by means of a Peirls substitution \cite{PhysRevA.95.023601,mccaul2020controlling,mccaul2020driven}, where $\hat{\Omega}\hat{F}$ may be removed from the equation of motion with the substitution $\hat{\Omega} \to \hat{\Omega} {\rm e}^{-i\frac{1}{\hbar}\hat{F}t}$. From this perspective it is clear that $\hat{F}$ may therefore be interpreted as the generator of the phase of the non self-adjoint waveoperator. From this perspective, the requirement that $\hat{F}$ be self-adjoint (i.e. Hermitian) is necessary to preserve the relationship given by Eq.\eqref{EqRhoThroughOmega}.}

{The existence of $\hat{F}$ does however permit some conceptual freedom in how to interpet the evolution of $\hat{\Omega}$. For instance the case of $\hat{F} = 0$ of the waveoperator description has been studied in \cite{uhlmann_parallel_1986, tronci_momentum_2019}, and produces an identical Liouville equation of motion as for $\hat{\rho}$. Furthermore, in such a case it is clear that for pure states, we have $\hat{\rho}=\hat{\Omega}$, while a non-zero $\hat{F}$ will introduce for pure states a phase relationship $\hat{\rho}={\rm e}^{i\frac{1}{\hbar}\hat{F}t}\hat{\Omega}$.   If one were to choose $\hat{F}=\hat{H}$, then the dynamics of $\hat{\Omega}$ would be Schr{\"o}dinger-like, but still reproduce the Liouville equation for $\hat{\rho}$. This translates into a small numerical advantage, insofar as only $\hat{H}\hat{\Omega}$ need be calculated, rather than the full commutator. Finally, a more concrete exploitation of this freedom may be found in \cite{ilya}, noting that at each time step $\hat{F}$ may be chosen so as to maintain a lower triangular shape for $\hat{\Omega}$, and thus minimise the number of coefficients that must be propagated. Being able to $\hat{\Omega}$ in this lower-triangular form is broadly applicable, as evidenced by its use in machine-learning assisted quantum state estimation \cite{Lohani2020}. Fundamentally however, the presence of $\hat{F}$ cannot change the character of observable dynamics, and the main motivation for its inclusion in the present article is for the sake of completeness.}

{The second and most important benefit} of the waveoperator formalism is conceptual. Specifically we shall see that when employed in concert with the technique of canonical purification, we obtain both a direct correspondence to the Wigner phase function, as well as a generally applicable procedure for taking the classical limit of a quantum system. {Furthermore, when considering dissipative and imaginary-time evolutions, we find that the waveoperator representation can possess markedly different properties as compared to the dynamics of $\hat{\rho}$. Finally, the guarantee of positivity and ability to consistently perform classical limits are two properties which may allow the waveoperator to serve as the basis} for a physically consistent model of a quantum-classical hybrid, but in the present work we restrict ourselves to the context of a closed system, where we are able to demonstrate the aforementioned classical limit. 

\section{Canonical purification of the waveoperator \label{Sec:purification}}

In this section we will establish a close link between the proposed waveoperator description of quantum mechanics and the notion of purification in quantum information theory (see chapter 5 in \cite{wilde_quantum_2017}). Expressing the waveoperator in a purified form will later allow for the introduction of Bopp operators, and the establishment of a classical limit for the formalism. To perform the purification, we first choose an arbitrary orthogonal time-independent basis $\{ \ket{k} \} \subset \cal{H}$ in a Hilbert space $\cal{H}$. This allows us to define a mapping from an operator $\hat{\Omega}$ acting on $\cal{H}$  to a vector $\ket{\hat{\Omega}} \in \cal{H} \otimes \cal{H}$ as 
\begin{equation}\label{EqVectorizationDefinition}
     \ket{\hat{\Omega}} \equiv \sum_k \hat{\Omega} \ket{k} \otimes \ket{k} = (\hat{\Omega} \otimes \hat{1}) \ket{\omega},
\end{equation}
where
\begin{equation}
    \ket{\omega}=\sum_k \ket{k}\otimes\ket{k}. 
\end{equation}
The transformation given by Eq.~(\ref{EqVectorizationDefinition}) is closely related to the the concept of canonical purification (see page 166 of \cite{wilde_quantum_2017}), while in linear algebra, the mapping is also known as row-major vectorization. Since Eq.~(\ref{EqVectorizationDefinition}) is a purification of the density matrix $\hat{\rho}$, the latter can be recovered as a partial trace,
\begin{equation}
    \hat{\rho} = \Tr' \ket{\hat{\Omega}}\bra{\hat{\Omega}} 
    \equiv \sum_{k} (\hat{1} \otimes \bra{k}) \ket{\hat{\Omega}}\bra{\hat{\Omega}}
    (\hat{1} \otimes \ket{k}).
\end{equation}
A number of important identities can be derived from the definition of Eq.~(\ref{EqVectorizationDefinition})
\begin{subequations}\label{EqVectorizationIdentities}
\begin{eqnarray}
    \ket{\hat{A}\hat{\Omega}} = (\hat{A} \otimes \hat{1}) \ket{\hat{\Omega}}, \label{EqVectorization1} \\
    \braket{\hat{A}|\hat{B}} = \Tr (\hat{A}^{\dagger} \hat{B}), \label{EqVectorization3} \\
    \ket{\hat{\Omega}\hat{A}} = (\hat{1} \otimes \hat{A}^T) \ket{\hat{\Omega}},  \label{EqVectorization2}
\end{eqnarray}
\end{subequations}
where $\hat{A}^T$ denotes the transpose of $\hat{A}$.
The latter identity  Eq.~\eqref{EqVectorization2} is a consequence of the following ``ricochet'' property:
\begin{eqnarray}
     \hat{A}\otimes\hat{1} \ket{\omega}&=\sum_{ijk}a_{ij}\ket{i}\braket{j|k}\otimes\ket{k} &=\sum_{ijk}a_{ij}\delta_{jk}\ket{i}\otimes\ket{k} \notag\\
    & =\sum_{ik}a_{ik}\ket{i}\otimes\ket{k}&=\sum_{ij} \ket{k}\otimes a_{ki}\ket{i} \notag\\
     & =\sum_{ijk}\ket{k}\otimes a_{ji}\delta_{jk}\ket{i}&= \sum_{ijk}\ket{k}\otimes a_{ji}\ket{i}\braket{j|k} \notag\\
     &=\hat{1}\otimes\hat{A}^T\ket{\omega}.&
\end{eqnarray}
When this is combined with the fact that any operators of the form $\hat{1}\otimes\hat{A}$ and $\hat{B}\otimes\hat{1}$ will commute, we obtain Eq.~\eqref{EqVectorization2}.

By combining Eq.~\eqref{EqWaveOperatorDynamics} with Eq.~\eqref{EqVectorizationIdentities}, it is possible to express the evolution of the waveoperator state in a  Schr\"odinger-like form
\begin{eqnarray}
    i\hbar\partial_t \ket{\Omega} &=& \left( \hat{H}\otimes \hat{1} - \hat{1} \otimes(\hat{H} + \hat{F})^T \right) \ket{\Omega}, \label{EqWaveOperSchrodinger}\\
    \braket{O} &=& \braket{\Omega| \hat{O} \otimes \hat{1}|\Omega}. \label{EqOAverageOmega}
\end{eqnarray}
The free choice of $\hat{F}$ also means that this evolution can correspond either to a Liouville-type commutator evolution when $\hat{F}=0$, or a Schr\"odinger equation with an ancillary space when $\hat{F}=-\hat{H}$. 
The dynamics denoted by Eq.~(\ref{EqWaveOperSchrodinger}) can also be arrived at from a Dirac-Frankel variational principle \cite{RAAB2000674},
\begin{equation}
    \delta \Re \int_{t_i}^{t_f} \braket{\Omega(t)| i\hbar\partial_t - \left( \hat{H}\otimes \hat{1} - \hat{1} \otimes(\hat{H} + \hat{F})^T \right) |\Omega(t)}dt = 0,
\end{equation}
where the choice of the ``phase generator'' $\hat{F}$ in (\ref{EqWaveOperSchrodinger}) does not affect the values of the observables since
\begin{eqnarray}\label{EqEhrenfestTheoremsWaveOpt}
    i\hbar \partial_t \braket{O} = \bra{\Omega} [ \hat{O}, \hat{H} ] \otimes \hat{1} \ket{\Omega}. 
\end{eqnarray}

The choice of an orthonormal basis in Eq.~(\ref{EqVectorizationDefinition}) to construct the purification of the waveoperator is equivalent to fixing the ``phase generator'' $\hat{F}$, and  hence bears no observational consequences. If $\ket{\Omega}$ and $\ket{\Omega'}$ denote two purifications of $\hat{\Omega}$ corresponding to the different bases $\{ \ket{k} \}$ and $\{ \ket{k'} \}$, then there exists a a unitary $\hat{U}$ such that $\ket{\Omega} = (\hat{1} \otimes \hat{U}) \ket{\Omega'}$  \cite{wilde_quantum_2017}. Then, Eq.~(\ref{EqWaveOperSchrodinger}) is invariant under the ``gauge'' transformation 
\begin{equation}
    \ket{\Omega} \to \ket{\Omega'}, \qquad
    \hat{F} \to \left( \hat{U}^{\dagger} (\hat{H} + \hat{F})^T\hat{U} + \hat{G} \right)^T - \hat{H},
\end{equation}
where the self-adjoint $\hat{G}$ is defined as $i\hbar\partial_t \hat{U} =  \hat{G}\hat{U}$ (i.e. Stone's theorem) \cite{Stonetheorem}.

\section{Bopp operators for purified waveoperators \label{SecBoppOp}}
Having defined the waveoperator and its dynamics when represented as a purified state, we now show that this formalism provides a transparent method for the introduction of Bopp operators \cite{AIHP_1956__15_2_81_0, hillery_distribution_1984, Zueco_2007}. These not only allow one to transit between Hilbert and phase space representations of a quantum system, but also enable a classical limit to be taken transparently, as we shall find in a later section. For simplicity, hereafter we will consider system with one degree of freedom, but the extension to multidimensional case is trivial.

In anticipation of later developments and following the conventions of~\cite{cabrera_efficient_2015, bondar_efficient_2016, ciric_exponential_2023}, we shall refer to quantum coordinate and momentum variables as  $\hat{\bi{x}}$ and $\hat{\bi{p}}$, where the bold font is used to indicate their status as non-commuting quantum operators, rather than vectorial notation.  These will obey the Heisenberg canonical commutation relation
\begin{equation}
    [\hat{\bi{x}}, \hat{\bi{p}}] = i\hbar.
\end{equation}
We will also assume that the operator functions $H(\hat{\bi{x}}, \hat{\bi{p}})$ and $F(\hat{\bi{x}}, \hat{\bi{p}})$ are represented in a Weyl-symmetrized form. We then introduce \emph{the Bopp operators} as
\begin{eqnarray}
    \hat{x} = \frac{1}{2} \left( \hat{1} \otimes \hat{\bi{x}}^T + \hat{\bi{x}} \otimes \hat{1} \right), \quad
    \hat{p} = \frac{1}{2} \left( \hat{\bi{p}} \otimes \hat{1} + \hat{1} \otimes \hat{\bi{p}}^T \right), \nonumber\\
    \hat{\theta} = \frac{1}{\hbar}\left(  \hat{1} \otimes \hat{\bi{x}}^T - \hat{\bi{x}} \otimes \hat{1} \right),
    \quad
    \hat{\lambda} = \frac{1}{\hbar} \left( \hat{\bi{p}} \otimes \hat{1} - \hat{1} \otimes \hat{\bi{p}}^T \right)\ \label{EqBoppOpDef}
\end{eqnarray}
while the inverse transformations read
\begin{eqnarray}
    \hat{\bi{x}} \otimes \hat{1}  = \hat{x} - \frac{\hbar}{2} \hat{\theta}, \qquad 
     &\hat{\bi{p}} \otimes \hat{1} = \hat{p} + \frac{\hbar}{2} \hat{\lambda}, \nonumber\\
     \hat{1} \otimes \hat{\bi{x}}^T = \hat{x} + \frac{\hbar}{2} \hat{\theta}, \qquad 
     &\hat{1} \otimes \hat{\bi{p}}^T = \hat{p} - \frac{\hbar}{2} \hat{\lambda}.
     \label{EqBoppOpInverse}
\end{eqnarray}
The commutation relations of these Bopp operators can be calculated as (for example):
\begin{align}
     [\hat{x}, \hat{p}]= &\frac{1}{4}\left([\hat{\bi{x}},\hat{\bi{p}}]\otimes\hat{1}+\hat{1}\otimes[\hat{\bi{x}}^T,\hat{\bi{p}}^T]\right), \\
     [\hat{\theta}, \hat{\lambda}]= &-\frac{1}{2\hbar}\left([\hat{\bi{x}},\hat{\bi{p}}]\otimes\hat{1}+\hat{1}\otimes[\hat{\bi{x}}^T,\hat{\bi{p}}^T]\right).
\end{align}
Since $(\hat{\bi{x}}\hat{\bi{p}})^T = \hat{\bi{p}}^T \hat{\bi{x}}^T$, transposing the fundamental commutation relation yields the identity $[\hat{\bi{x}}^T,\hat{\bi{p}}^T]=-i\hbar$, and means the Bopp operators obey the following alegbra: 
\begin{subequations}\label{EqBoppOpAlgebra}
\begin{gather}\label{EqBoppOpAlgebraXPCommutator}
    [\hat{x}, \hat{p}] = [\hat{\theta}, \hat{\lambda}] = 0,\\
    [\hat{p}, \hat{\theta}] = [\hat{x}, \hat{\lambda}] = i.
\end{gather}
\end{subequations}
With the help of the identities  
\begin{eqnarray}
 & \hat{1} \otimes H(\hat{\bi{x}}^T, \hat{\bi{p}}^T) = H(\hat{1} \otimes \hat{\bi{x}}^T, \hat{1} \otimes  \hat{\bi{p}}^T),  \\
  &H(\hat{\bi{x}}, \hat{\bi{p}}) \otimes \hat{1} = H(\hat{\bi{x}} \otimes \hat{1}, \hat{\bi{p}} \otimes \hat{1}),
\end{eqnarray}
which are valid for any Weyl-symmetrized $\hat H$, the equations for the state dynamics and expectations read:
\begin{align}
    \label{EqBoppevolution}i\hbar\partial_t \ket{\Omega} =& \hat G \ket{\Omega}, \\
    \label{EqForG} \hat G =& H(\hat{x} - \frac{\hbar}{2} \hat{\theta}, \hat{p} + \frac{\hbar}{2} \hat{\lambda})
    - H(\hat{x} + \frac{\hbar}{2} \hat{\theta}, \hat{p} - \frac{\hbar}{2} \hat{\lambda}) \notag \\ &- F(\hat{x} + \frac{\hbar}{2} \hat{\theta}, \hat{p} - \frac{\hbar}{2} \hat{\lambda}),\\
    \label{EqBraKetOBopp}
    \braket{O} =& \braket{\Omega|  O(\hat{x} - \frac{\hbar}{2} \hat{\theta}, \hat{p} + \frac{\hbar}{2} \hat{\lambda}) |\Omega}. 
\end{align}
We note that Eqs.~(\ref{EqBoppevolution})-(\ref{EqBraKetOBopp}) have been derived in Ref.~\cite{bondar_operational_2012}, but from an entirely different perspective.

Since $\hat{x}$ and $\hat{p}$ commute, they share a common eigenbasis
\begin{align}\label{EqXPRepresentation}
    \hat{x}\ket{xp} &= x\ket{xp}, \qquad \hat{p}\ket{xp} = p\ket{xp}, \\ &\hat{1} \otimes \hat{1} = \int dxdp \ket{xp}\bra{xp}.
\end{align}
It follows from the commutator relationship~(\ref{EqBoppOpAlgebra}) that 
\begin{align}
    &\braket{xp|\hat{x}|\Omega} = x \braket{xp|\Omega}, \quad
    \braket{xp|\hat{\lambda}|\Omega} = -i\partial_x \braket{xp|\Omega}, \notag \\
    &\braket{xp|\hat{p}|\Omega} = p \braket{xp|\Omega},  \quad
    \braket{xp|\hat{\theta}|\Omega} = -i\partial_p \braket{xp|\Omega}.
\end{align}
Hence,
\begin{align}
     i\hbar\partial_t \braket{xp|\Omega} =& \Big( H(x_+, p_-) - H(x_-, p_+ ) \nonumber\\
 &-  F(x_-, p_+)
    \Big) \braket{xp|\Omega}, \label{EqKetOBoppXPRepr} \\
    \braket{O} =& \int dxdp \braket{\Omega|xp} O(x_+,p_-) \braket{xp|\Omega}, \label{EqAverageOXPRepr} \\
    x_{\pm}=&x \pm i\frac{\hbar}{2} \partial_p, \ \ p_{\pm}=  p \pm i \frac{\hbar}{2} \partial_x.
\end{align}
When $F=0$ Eq.~(\ref{EqKetOBoppXPRepr}) coincides with the equation of motion for the Wigner function (see, e.g., Eq.~(2.77) in \cite{hillery_distribution_1984}). In this case however, the original waveoperator is not restricted to representing a pure state, meaning that Eq.~\eqref{EqBraKetOBopp} in cojunction with Eq.~(\ref{EqBoppevolution}) extend the interpretation of the Wigner function as a wave function \cite{bondar_wigner_2013} to include the general case of mixed states. 

\section{The phase-space representation of the waveoperator \label{Sec:phasespace}}

In this section, we will provide an alternative derivation of Eq.~(\ref{EqKetOBoppXPRepr}) and Eq.~(\ref{EqAverageOXPRepr}). The Wigner-Weyl transformation of equations (\ref{EqWaveOperatorDynamics}) and (\ref{EqOAverageWaveOperator}) read
\begin{align}
    i\hbar\partial_t \Omega(x,p) =& H(x,p) \star \Omega(x, p) - \Omega(x,p) \star H(x, p) \notag \\ &-   \Omega(x,p) \star F(x, p), \\
    \braket{O} =&\int dxdp\Omega(x,p)^* \star O(x, p) \star \Omega(x, p),
\end{align}
where $\star$ denotes the Moyal product, $H(x,p)$, $\Omega(x,p)$, $F(x,p)$, and $O(x,p)$ are the Weyl symbols for the operators $\hat{H}$, $\hat{\Omega}$, $\hat{F}$, and $\hat{O}$, respectively.
Utilizing the ``lone star'' identity $\int  f(x,p) \star g(x,p) dxdp = \int  f(x,p)g(x,p) dxdp$ (see equation~(16) in \cite{curtright_concise_2014}) and
\begin{eqnarray}
    f(x,p) \star g(x, p) = f\left(x_+, p_- \right) g(x,p), \\
    g(x,p) \star f(x, p) = f\left(x_-, p_+\right) g(x,p),
\end{eqnarray}
(see, e.g., equations (12) and (13) in \cite{curtright_concise_2014, cabrera_efficient_2015}),
we obtain
\begin{align}
    i\hbar\partial_t \Omega(x,p) =& \Big( H(x_+, p_- )
     - H(x_-, p_+ ) \nonumber\\
    &- F(x_-,p_+)
    \Big) \Omega(x,p), \label{MasterWignerWeylEqForWaveOperator}\\
    \braket{O} =& \int dxdp \, \Omega(x,p)^* O\left(x_+,p_- \right)  \Omega(x, p). \label{EqAverageOWignerOmega}
\end{align}
Comparing these two equations with Eq.(\ref{EqKetOBoppXPRepr}) and Eq.(\ref{EqAverageOXPRepr}), we conclude that $\braket{xp|\Omega} \equiv \Omega(x,p)$, i.e., $\braket{xp|\Omega}$ is the Wigner-Weyl transform of $\hat{\Omega}$.

We can also recover a more direct interpretation of $\Omega(x,p)$ in the case that $W(x,p)$ is the Wigner function for a pure quantum state $\hat{\rho}$. Recalling that purity implies $W(x,p) \star W(x,p) = \frac{1}{2\pi\hbar} W(x,p)$ (see, e.g., equation (25) in \cite{curtright_concise_2014}), one shows
\begin{eqnarray}
    \braket{O} &= \int dxdp\, O(x,p) W(x,p) = \int dxdp\, O(x,p) \star W(x,p) \nonumber\\
            &= 2\pi\hbar \int dxdp\, O(x,p) \star W(x,p) \star W(x,p) \nonumber\\
            &= 2\pi\hbar \int dxdp\, W(x,p) O(x,p) \star W(x,p) \nonumber\\
            &= 2\pi\hbar \int dxdp\, W(x,p) O\left(x_+,p_-\right) W(x,p). \label{EqWignerAsAWaveFuncAver}
\end{eqnarray}
Since the Wigner function is real by construction, Eq.~(\ref{EqWignerAsAWaveFuncAver}) is recovered from Eq.~(\ref{EqAverageOWignerOmega}) and Eq.~\eqref{MasterWignerWeylEqForWaveOperator} in the case $F{=}0$ if
\begin{equation}\label{EqWignerAsAWaveFunc}
    \Omega(x,p) = \sqrt{2\pi\hbar} W(x,p).
\end{equation}
Eq.~(\ref{EqWignerAsAWaveFuncAver}) and Eq.~\eqref{EqWignerAsAWaveFunc} therefore provide an alternative and much more simple derivation of the interpretation, put forth in \cite{bondar_wigner_2013}, of the Wigner function for a pure quantum system as a Koopman–von Neumann wave function. In particular Eqs.~(\ref{EqWignerAsAWaveFuncAver}) and (\ref{EqWignerAsAWaveFunc}) coincide with  Eqs.~(19) and (8) in \cite{bondar_wigner_2013}. In the general mixed case, we are still able to identify the waveoperator with the Wigner function thanks to Eqs.~(\ref{EqKetOBoppXPRepr}) and (\ref{EqAverageOXPRepr}).

\section{The classical limit of the waveoperator description \label{SecClassicalLimit}}
The proposed formalism also offers a direct route to the classical limit of quantum dynamics, where the Koopman-von Neumann representation of classical dynamics is naturally recovered. Beginning from Eq.~\eqref{EqBoppevolution}, we first scale our arbitrary phase $F\to \hbar F$, purely as a matter of convenience when taking the classical limit. Having done so, we then Taylor expand the Hamiltonian around the Bopp operators:
\begin{align}
 H(\hat{x} \mp \frac{\hbar}{2} \hat{\theta}, \hat{p} \pm \frac{\hbar}{2} \hat{\lambda})=&H(\hat{x},\hat{p}) \pm \frac{\hbar}{2} \partial_pH(\hat{x},\hat{p}) \hat{\lambda} \notag \\ \label{EqHamexpansion}  &\mp  \frac{\hbar}{2} \partial_xH(\hat{x},\hat{p}) \hat{\theta} +O(\hbar^2).
\end{align}
Inserting this into Eq.~\eqref{EqBoppevolution} and Eq.~\eqref{EqForG} we obtain
\begin{equation}
    \hat{G} = \partial_pH(\hat{x},\hat{p}) \hat{\lambda} -  \partial_xH(\hat{x},\hat{p}) \hat{\theta} +  F(\hat{x} + \frac{\hbar}{2} \hat{\theta}, \hat{p} - \frac{\hbar}{2} \hat{\lambda})+O(\hbar^2).
\end{equation}
Taking $\lim_{\hbar\to0}\hat{G}$ recovers the well-known KvN propagator, describing classical dynamics:
\begin{equation}
\label{Eq:ClassicalLimit}
    i\partial_t \ket{\Omega} =\left[\partial_pH(\hat{x},\hat{p}) \hat{\lambda} -  \partial_xH(\hat{x},\hat{p}) \hat{\theta} +  F(\hat{x}, \hat{p})\right]\ket{\Omega}. 
\end{equation}
We see that the phase generator in the classical limit corresponds to the arbitrary phase-space function one obtains in standard derivations of  KvN \cite{PhysRevE.99.062121,McCaul_2022,mezic1,mezic2,mezic3}, which itself relates KvN to alternative dynamical equations such as Koopman-van Hove (KvH) \cite{Bondar2019,Kirillov2001,doi:10.1142/4721}. 

The connection between Eq.~\eqref{Eq:ClassicalLimit} and the standard Liouville equation for the density can be made explicit by expressing this equation of motion in phase space:
\begin{equation}
    \partial_t \Omega(x,p)= \left\{H(x,p), \Omega(x,p)\right\} -iF(x,p)\Omega(x,p),
\end{equation}
where $\left\{\cdot,\cdot\right\}$ indicates the Poisson bracket. Using $\rho(x,p)=\left|\Omega(x,p)\right|^2$, we immediately recover the Liouville equation for the density. 

 It is also interesting to note that when expanding the right hand sides of Eqs.~\eqref{EqBoppevolution} and \eqref{EqHamexpansion} in series in $\hbar$ all terms corresponding to even powers of $\hbar$ will cancel. An immediate consequence is that for quadratic Hamiltonians, the $\hbar\to 0$ limit may only affect arbitrary phase term $F$. Otherwise, the waveoperator and Wigner function formalisms share the same property that the quantum equations of motion for quadratic systems remain unchanged in the classical limit $\hbar\to0$. 

\section{waveoperator representation of thermal states \label{Sec:Imaginarytime}}
The waveoperator formalism is also instructive when considering the quantum correction to equilibrium states. Recall that the density matrix for the Gibbs state at temperature $k_B T{=}\tfrac1{\beta}$ can be found (up to normalisation) via imaginary time propagation starting from $\beta{=}0$ (see, e.g., \cite{DMM, bondar_efficient_2016}):
\begin{equation}
\label{EqBlochevolution}
\partial_\beta\hat{\rho}=-\frac{1}{2}\left(\hat{H}\hat{\rho}+\hat{\rho}\hat{H}\right), \ \ \hat{\rho}(0)=\hat{1}.
\end{equation}
The solution to this equation is clearly the {unnormalised density} $\hat{\rho}(\beta)={\rm e}^{-\beta\hat{H}}$, which selects the ground state as $\beta \to \infty$. The matching equation for the waveoperator is:
{
\begin{equation}
\partial_\beta \hat{\Omega}=-\frac{1}{4}\left(\hat{H}\hat{\Omega}+\hat{\Omega}\hat{H}\right), \ \ \hat{\Omega}(0)=\hat{1}. \label{eq:waveoperatorBloch}  
\end{equation}
}
Eq.~\eqref{eq:waveoperatorBloch} can be proved directly by showing that  the density matrix $\hat{\rho}=\hat{\Omega}\hat{\Omega}^\dagger$ is the solution to Eq.~\eqref{EqBlochevolution} when $\hat{\Omega}$ solves Eq.~\eqref{eq:waveoperatorBloch}. {Significantly, while one could introduce a phase generator into this equation for $\hat{\Omega}$ using $\hat{H}\to\hat{H}+i\hat{F}$, in this case $\hat{F}$ must not only be self-adjoint, but commute with $\hat{H}$ in order to reproduce Eq.~\eqref{EqBlochevolution}. Such a restriction means that the potentially exploitable freedom of choice for $\hat{F}$ in real-time does not exist in this instance, and we therefore set $\hat{F}=0$.}

By vectorizing the thermal state waveoperator $\hat\Omega$ according to Eq.~\eqref{EqVectorizationDefinition}, Eq.~\eqref{eq:waveoperatorBloch} can be restated in terms of Bopp operators as
\begin{equation}
\label{eq:waveoperatorBlochVectorized}
  \partial_\beta \ket{\Omega} =-\frac{1}{4}\left[H(\hat{x} - \frac{\hbar}{2} \hat{\theta}, \hat{p} + \frac{\hbar}{2} \hat{\lambda})
    + H(\hat{x} + \frac{\hbar}{2} \hat{\theta}, \hat{p} - \frac{\hbar}{2} \hat{\lambda})\right]\ket{\Omega}.   
\end{equation}
Series expansion of the right hand side of Eq.~\eqref{eq:waveoperatorBlochVectorized} in $\hbar$ gives
\begin{align}
  \partial_\beta \ket{\Omega} &=\frac{1}{2}H(\hat{x},\hat{p})\ket{\Omega}\notag \\ \label{EqImaginaryWaveOperatorEvolution} +&\frac{\hbar^2}{4}\left(\partial^2_x H(\hat{x},\hat{p})\hat{\theta}^2 +\partial^2_p H(\hat{x},\hat{p})\hat{\lambda}^2\right)\ket{\Omega}
+O(\hbar^4).   
\end{align}
Thus, the lowest order quantum correction to the ground or thermal state is of order $\hbar^2$, and only the terms corresponding to even powers of $\hbar$ survive. This means that unlike in real time, Eq.~\eqref{eq:waveoperatorBlochVectorized} retains its form in the classical limit $\hbar\to0$ only for \textit{linear} Hamiltonians. It is interesting to observe that the semiclassical correction has the form of a Fokker-Planck like diffusive term when expressed in phase space. {We note also that imaginary time evolution remains a non-trace preserving operation regardless of whether one represents it with $\hat{\rho}$ or $\hat{\Omega}$, and the unnormalised states presented here will still require the calculation of a partition function. Nevertheless, being able to represent the waveoperator semi-classical dynamics as the Fokker-Planck equation in phase space means that regardless of the system dimension, normalisation can be achieved with summation over a discretised grid of phase-space coordinates, rather than tracing over a Hilbert space whose dimensionality has an exponential dependence on system size.} 

\begin{figure}[ht!]
\captionsetup{width=0.48\textwidth}
\centering
\begin{center}
  \includegraphics[width=0.5\textwidth]{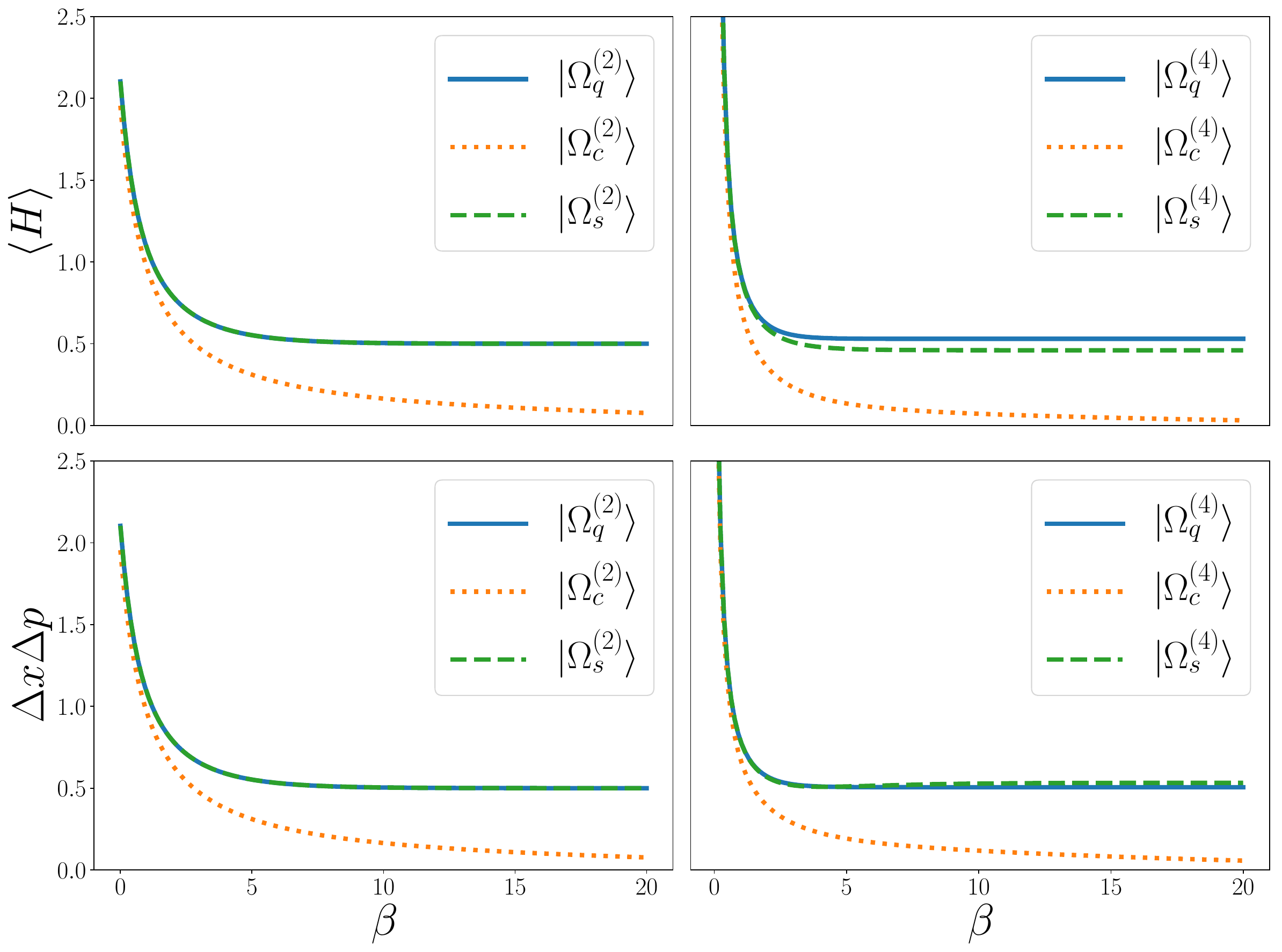}
  \end{center}
  \begin{center}
  \caption{Comparison of the imaginary time dynamics for quantum, semiclassical and classical systems, in the cases of quadratic and quartic Hamiltonians. \added{All quantities are given in natural units using $\hbar = k_B =1$.} In both cases, the expected energy of the system is distinguished by the zero-point energy present in the quantum and semiclassical evolutions. Finally, the uncertainty relation in position and momentum is clearly respected in the quantum and semiclassical system, while in the classical system it approaches zero, reflecting the zero energy classical ground state.}
  \label{fig:quadratic}
 \end{center}
\end{figure}

In order to showcase the distinction between classical and quantum worlds within the waveoperator formalism, let us compare the thermal states for benchmark one-dimensional quadratic and quartic systems. These will be described by the Hamiltonians
\begin{equation}
\hat{H}^{(n)}=\frac{1}{2} \bi{\hat{p}}^2 +\frac{1}{2}\bi{\hat{x}}^n, 
\end{equation}
where $n=2$ and $n=4$, respectively. Let us consider three levels of approximation to Eq.~\eqref{eq:waveoperatorBlochVectorized}: We label $\ket{\Omega^{(n)}_q}$ as the state obtained when evolving using the fully quantum Eq.~\eqref{eq:waveoperatorBloch}. A semiclassical $\ket{\Omega^{(n)}_s}$ is derived from Eq.~\eqref{EqImaginaryWaveOperatorEvolution} by dropping the $O(\hbar^4)$ terms, and finally $\ket{\Omega^{(n)}_c}$ is the evolution using the $\hbar \to 0$ limit of Eq.~\eqref{EqImaginaryWaveOperatorEvolution}, which additionally wipes out $O(\hbar^2)$ terms. Fig. \ref{fig:quadratic} illustrates these three types of evolution. As expected, the quantum and semiclassical evolutions are identical for an $n=2$ quadratic Hamiltonian, but surprisingly produce only slightly different results for the quartic $n=4$ Hamiltonian. In all cases however, the distinction between classical and quantum evolutions is clear, where in the former case the absence of a fundamental commutation relation (and therefore a zero-point energy) is reflected in both the ground state energy and $\Delta x\Delta p$, as shown in  Fig. \ref{fig:quadratic}.

\section{Non-linearity in waveoperator dynamics \label{sec:nonlinearity}}

{ There exist a number of phenomena whose description incorporates non-linear models, such as Bose-Einstein condensates \cite{Lindberg_2023}, or wavefunction collapse \cite{PhysRevA.47.3538}. A natural question to ask is therefore whether the waveoperator and density matrix representations may differ in the degree of non-linearity of their dynamics. There are in fact a number of scenarios in which a linear evolution in either $\hat{\rho}$ of $\hat{\Omega}$ leads to a \textit{non-linear} evolution in the complementary representation. Here we provide two conjugate examples of this phenomenon. 

First, let us consider a thermal state $\hat{\rho}(\beta)=\frac{1}{Z}{\rm e}^{-\beta\hat{H}}$. Representing this state as a waveoperator, we find
\begin{equation}
    \ket{\Omega(\beta)}=\frac{1}{\sqrt{Z}}\sum_j  {\rm e}^{-\frac{\beta}{2}\hat{H}}\ket{E_j} \otimes \ket{E_j},
\end{equation}
where $\ket{E_j}$ are eigenvectors of $\hat{H}$.
Remarkably, this object is one of some importance in holography \cite{Maldacena1999}, as it may represent a black hole in AdS conformal field theories \cite{Maldacena_2003}, and hence provide a platform for the study of many quantum phenomena in this context \cite{Almheiri2013, Susskind}. In such a context it is known as the \textit{thermofield double} (TFD) state, and questions of how to prepare this state are a focus of ongoing research \cite{PhysRevLett.127.080602}. One consequence of this has been the finding that for a harmonic oscillator of mass $m$ and frequency $\omega$, using
\begin{equation}
\hat{a}=\sqrt{\frac{m\omega}{2\hbar}}\left(\hat{\bi{x}} +i\frac{1}{m\omega}\hat{\bi{p}}\right),
\end{equation}
the TFD may be written as \cite{SciPostPhys.6.3.034}
\begin{equation}
    \ket{\Omega(\beta)}=\frac{1}{\sqrt{1-{\rm e}^{-\beta \omega}}}\sum_n  {\rm e}^{-\frac{1}{2}n\beta\omega}\ket{n} \otimes \ket{n}.
\end{equation}
where $\ket{n}$ is a state in the second-quantised basis. In this instance, it is possible to express this normalised waveoperator as the unitary evolution of a two-mode squeezing Hamiltonian on the vacuum state. This is given by \cite{SciPostPhys.6.3.034}:
\begin{equation}
\ket{\Omega(t)} = {\rm e}^{-\frac{i}{\hbar}t\left(\hat{\bi{x}}\otimes\hat{\bi{p}}+\hat{\bi{p}}\otimes\hat{\bi{x}} \right)} \ket{0}\otimes \ket{0},
\end{equation}
where 
\begin{equation}
t=\frac{1}{2}\log\left(\frac{1+{\rm e}^{-\frac{1}{2}\beta \omega}}{1-{\rm e}^{-\frac{1}{2}\beta \omega}}\right).
\end{equation}
Most interestingly, in the waveoperator framework this leads to a form of dynamics where (up to a phase) $\hat{\Omega}(0)=\hat{\rho}(0)=\ket{0}\bra{0}$, and the generator does not conform to the form specified by Eq.~\eqref{EqWaveOperSchrodinger}. This is due to the fact that this generator possesses non-trivial operators on both sides of the tensor product. Nevertheless, Eqs.~(\ref{EqVectorization1}-\ref{EqVectorization2}) may be employed to depurify $\ket{\Omega(t)}$, leading to a \textit{real-time} equation of motion describing thermalisation on the level of $\hat{\Omega}$ in the number basis:
\begin{equation}
i \hbar \partial_t \hat{\Omega}= \hat{\bi{p}}\hat{\Omega}\hat{\bi{x}}^T+\hat{\bi{x}}\hat{\Omega}\hat{\bi{p}}^T.
\end{equation}
The sandwiching of $\hat{\Omega}$ between operators has important consequences for the corresponding dynamics of $\hat{\rho}$, namely that it becomes intrinsically \textit{non-linear}. To demonstrate this we first define the similarity transformation
\added{
\begin{equation}
    \bar{A}(\hat{\Omega})\equiv\bar{A}= \hat{\Omega}\hat{A}^T \hat{\Omega}^{-1} .
\end{equation}
Note that for any finite temperature, the eigenvalues of $\hat{\Omega}$ will be $\{{\rm e}^{-\frac{\beta}{2}E_j}\}$. Given these are all non-zero, $\hat{\Omega}^{-1}$ is guaranteed to exist.} The evolution of a thermal state in real-time for the density may then be described with:
\begin{equation}
    i\hbar \partial_t \hat{\rho}=(\hat{\bi x}\bar{ \bi p} +\hat{ \bi p}\bar{ \bi x})\hat{\rho}-\hat{\rho}(\bar{\bi p}^\dagger \hat{ \bi x}\ +\bar{\bi x}^\dagger\hat{\bi p})
\end{equation}
revealing that such a thermalising evolution has drastically different properties depending on the representation one chooses for the state. 

There also exists a complementary scenario in which the non-linearity relationship between the two representations is reversed. If we consider a Lindblad equation of the form \cite{McCaulensemble}
\begin{equation}
 \partial_t \hat{\rho}=-\frac{i}{\hbar} \left[\hat{H},\hat{\rho}\right] + 2\hat{L}\hat{\rho}\hat{L}^\dagger -\left\{\hat{L}^\dagger \hat{L}, \hat{\rho} \right\},
\end{equation}
we find that the structure of the $\hat{L}\hat{\rho}\hat{L}^\dagger$ term leads to a dissipative equation for $\hat{\Omega}$.  Using a slightly modified definition $\bar{L}=\hat{\Omega}^{-1}\hat{L} \hat{\Omega}$, we obtain \cite{ilya}
\added{
\begin{equation}
\partial_t \hat{\Omega} = -\frac{i}{\hbar} \left[\hat{H},\hat{\Omega}\right] +\hat{L}\hat{\Omega}\bar{L}^\dagger -\hat{L}^\dagger \hat{L}\hat{\Omega} 
\end{equation}
}
where once again a linear equation has been transformed into a non-linear one due to the presence of an operator-state-operator sandwich term in the linear dynamics.} 


\section{Discussion \label{Sec:Discussion}}
Here we have presented a novel representation of Hilbert space dynamics, where positivity is automatically preserved by performing dynamics on the square root of the density operator. 
One advantage of the present formalism is that it may be set in Hilbert space with a Schr\"{o}dinger-like equation of motion. Consequently it is possible to use all of the highly efficient tools developed for these dynamics (e.g. tensor network algorithms) when performing waveoperator calculations. Furthermore, incorporating purification allows one to introduce the Bopp operators to this formalism. This has resulted in a phase space representation of the waveoperator, which in turn allows us to identify the Wigner function as the projection of the waveoperator onto phase space.

{Practical applications of the waveoperator formalism are already extant. For example, it has been used to formulate a method of thermal density matrix minimisation used to successfully calculate the equilibrium density matrix of a 6912 atom silicon cell \cite{DMM}. Furthermore, the structural advantages of the waveoperator representation have already inspired proposals for the simulation of Lindbladian dynamics in the context of quantum circuits \cite{Patel2023} as well as machine-learning assisted quantum state estimation \cite{Lohani2020}.}

Taking the classical limit of the waveoperator formalism, we find that it corresponds to the KvN representation of classical dynamics. For quadratic Hamiltonians, this correspondence to classical dynamics is exact even before taking any $\hbar\to0$ limit. This mirrors similar results to be found in the path integral and Wigner representations of dynamics for quadratic systems. In the former case, a saddle point approximation ensures only paths corresponding to the classical action contribute to the propagator, while the Moyal bracket operation evolving the quasiprobability $W(x,p)$ reduces to the Poisson bracket in the phase space representation.  

When performing an analogous procedure in imaginary time, an $O(\hbar^2)$ correction distinguishes quantum and classical quadratic systems, suggesting the most significant difference between the quantum and classical regime is in the ground state systems inhabit, rather than their real time dynamics (see \cite{kajari_inertial_2010}). {A potential application of this new semiclassical expansion is in the calculation of tunnelling rates \cite{PhysRevResearch.3.033019}, which at present employ semiclassical path-integrals. One future avenue of research is to investigate how this is related to the semiclassical waveoperator, and what if any advantage the latter may possess.}  The fact that the semiclassical expansion of the imaginary time evolution exhibits a quadratic correction to the classical Hamiltonian is strikingly similar to another context in which the equilibrium state of a system is determined by a correction to its bare Hamiltonian. Specifically, when considering a system strongly coupled to an environment, its thermal state is described by a ``Hamiltonian of mean force'' that accounts for the environmental interaction. In those cases where this effective Hamiltonian is known \cite{PhysRevA.106.012204,Cerisola,PhysRevB.95.125124,doi:10.1116/5.0073853,Miller2018}, the correction to the bare Hamiltonian is \textit{also} quadratic rather than linear. It is tempting to speculate that these two phenomena may be related to each other.

There are a number of potential extensions to this formalism. For instance, the introduction of commuting Bopp operators in previous sections relies on the canonical commutation relation in an infinite dimensional space of operators \cite{doi:10.1063/1.1665849}. Finite dimensional Hilbert spaces are more restrictive and generally do not allow introducing the analogs of the Bopp operators with the desired commutation properties. Nevertheless, one might employ (for example) a Jordan-Schwinger map \cite{Sakurai2010-mw} to represent such a finite dimensional system. This would introduce an oscillator basis obeying canonical commutation relations in the continuum limit, and open a route to calculations analogous to those presented here.  
More generally, the hunt for novel efficient representations for interacting systems is one of the chief motivations for the development of the waveoperator formalism. Specifically, the fact that positivity is automatically preserved is of vital importance when attempting to construct a hybrid formalism, where a partial classical limit is taken on one part of an interacting system. The importance of developing such formalisms should not be understated, given that all our interactions with the quantum world must be mediated through essentially classical devices, which will themselves have a quantum backreaction. The growing sophistication of quantum technology demands we be able to accurately describe such phenomena via a quantum-classical hybrid.  Traditionally however, such hybrid representations map to the quantum density operator $\hat{\rho}$, and the hybrid equations of motion derived (e.g. the Alexandrov-Gerasimenko-Kapral equation \cite{aleksandrov_statistical_1981, gerasimenko_dynamical_1982, kapral_mixed_1999}) do not necessarily preserve the positivity of the state \cite{Bondar2019}, calling into question the physicality of the dynamics. This remains an area of active research \cite{doi:10.1146/annurev.physchem.57.032905.104702,PhysRevD.37.3522,doi:10.1063/1.478811,doi:10.1063/1.481225, Cesare1, Cesare2}, with quantum computational approaches to this problem already being explored \cite{hybridcomputing}. It is our hope that the waveoperator formalism developed here will provide a useful tool in the ongoing development of hybrid system models.

\acknowledgments

G.M. and D.I.B. are supported by the U.S. Army Research Office (ARO) under grant W911NF-23-1-0288. The views and conclusions contained in this document are those of the authors and should not be interpreted as representing the official policies, either expressed or implied, of ARO, or the U.S. Government. The U.S. Government is authorized to reproduce and distribute reprints for Government purposes notwithstanding any copyright notation herein.




\end{document}